\documentclass[nofootinbib,superscriptaddress,a4paper,twocolumn]{revtex4-1}

\usepackage{geometry}
\usepackage[english]{babel}
\usepackage[latin1]{inputenc}
\usepackage{hyperref}
\usepackage{amsmath, amssymb, amsfonts, mathrsfs}
\usepackage{graphicx}

\graphicspath{{images/}}
\usepackage{xcolor}
\usepackage{exscale}
\usepackage{graphicx}
\usepackage{amsmath}
\usepackage{latexsym}
\usepackage{amsfonts}
\usepackage{amssymb}
\usepackage{bbm}
\usepackage{amsmath}
 
\usepackage{xspace}

\def\pmx{\begin{pmatrix}}
\def\emx{\end{pmatrix}}

\begin{document} 

\title{On Small Beams with Large Topological Charge II: \\ Photons, Electrons and Gravitational Waves}

\author{Mario Krenn}
\email{mario.krenn@univie.ac.at}
\affiliation{Vienna Center for Quantum Science \& Technology (VCQ), Faculty of Physics, University of Vienna, Boltzmanngasse 5, 1090 Vienna, Austria.}
\affiliation{Institute for Quantum Optics and Quantum Information (IQOQI), Austrian Academy of Sciences, Boltzmanngasse 3, 1090 Vienna, Austria.}
\author{Anton Zeilinger}
\email{anton.zeilinger@univie.ac.at}
\affiliation{Vienna Center for Quantum Science \& Technology (VCQ), Faculty of Physics, University of Vienna, Boltzmanngasse 5, 1090 Vienna, Austria.}
\affiliation{Institute for Quantum Optics and Quantum Information (IQOQI), Austrian Academy of Sciences, Boltzmanngasse 3, 1090 Vienna, Austria.}

\begin{abstract}
Beams of light with a large topological charge significantly change their spatial structure when they are focused strongly. Physically, it can be explained by an emerging electromagnetic field component in the direction of propagation, which is neglected in the simplified scalar wave picture in optics.  Here we ask: Is this a specific photonic behavior, or can similar phenomena also be predicted for other species of particles? We show that the same modification of the spatial structure exists for relativistic electrons as well as for focused gravitational waves. However, this is for different physical reasons: For electrons, which are described by the Dirac equation, the spatial structure changes due to a Spin-Orbit coupling in the relativistic regime. In gravitational waves described with linearized general relativity, the curvature of space-time between the transverse and propagation direction leads to the modification of the spatial structure. Thus, this universal phenomenon exists for both massive and massless elementary particles with Spin 1/2, 1 and 2. It would be very interesting whether other types of particles such as composite systems (neutrons or C$_{60}$) or neutrinos show a similar behaviour and how this phenomenon can be explained in a unified physical way.
\end{abstract}
\date{\today}
\maketitle

\section{Introduction}
Two years ago, we found in theoretical calculations that the intensity structure of light with large topological charge changes greatly when it is focused \cite{krenn2016small}. At that time, we asked whether this phenomenon also exists for other particle types, or whether light and matter waves differ fundamentally in the non-paraxial regime (since they are described by very different wave equations). Here we answer this question for electrons and gravitational waves: The same change of the intensity structure is theoretically predicted. The changes in intensity come from the interplay between different vector components of the fields. Surprisingly, for Maxwell fields, Dirac Spinors, and Riemann curvature tensors in linearized Gravity the same phenomenon happens, which points to a fundamental underlying reason that still remains to be discovered.

Now we first motivate the questions in general, and then show the predictions for photons, electrons and gravitational waves with the application of a single formalism.
\section{Background and Motivation}
\subsection{Paraxial Laguerre-Gauss modes}
The paraxial wave equation
\begin{eqnarray}
\left(\frac{\partial^2}{\partial x^2}+\frac{\partial^2}{\partial y^2}-2 i k \frac{\partial}{\partial z}\right)\psi(x,y,z)=0
\label{PWE1}
\end{eqnarray}
describes the propagation of beams of light which are only moderately focussed (more precisely, which have only small angles between their wave vector $k$ and the axis of propagation $z$, i.e. $k_{\perp}\ll k_z$, where $k_{\perp}$ and $k_z$ stand for the momentum in transverse direction and in the direction of propagation). Light beams can carry a discrete, large (theoretically unbounded) amount of orbital-angular momentum (OAM) or topological change \cite{allen1992orbital,erhard2018twisted}. Within the paraxial regime, such beams can be described as Laguerre-Gaussian modes in the following form
\begin{align}
LG&_{p,\ell}(r,\phi,z)=\nonumber\\
&=\sqrt{\frac{2p!}{\pi(p+\left | \ell \right |)!}}\frac{1}{w(z)}\left (\frac{\sqrt{2}r}{w(z)}\right)^{\left| \ell \right|}\nonumber\\
&\times L_{p}^{\left | \ell \right |} \left( \frac{2 r^{2}}{w(z)^{2}}\right) \exp\left(-\frac{r^{2}}{w(z)^{2}}\right)\nonumber\\
&\times \exp\left( -i\left( \frac{kr^{2}}{2R(z)} + \ell\phi-(2p+\left | \ell \right |+1)\varphi_{g} \right) \right)
\label{LGmode}
\end{align}
where $\ell$ stands for the orbital angular momentum mode number and $p$ is a radial mode number (which we set to $p=0$ for the rest of the article). $L_{p}^{\left| \ell \right|}$ are the Laguerre polynomials, $w(z)=w_{0}\sqrt{1+(\frac{z}{z_R})^{2}}$ is the gaussian beam waist at a distance $z$ from the focus, and a focal beam waist $w_0$. $z_{R}=\frac{\pi w_0^2}{\lambda}$ is the Rayleigh range, $R(z)=z\left(1+\left(\frac{z_R}{z}\right)^{2}\right)$ is the radius of curvature, $\lambda$ is the wavelength and $k=\frac{2\pi}{\lambda}$ is the wave number, and $\varphi_{g}=\arctan(\frac{z}{z_{R}})$ denotes the Gouy phase.

\subsection{Radial Scaling of LG modes}
Now we consider the radial scaling of LG modes, which -- in a naive treatment -- leads to an unphysical prediction. This can easily be calculated by using eq.(\ref{LGmode}) and finding the extreme point of the intensity distribution (for p=0 and z=0)
\begin{align}
\partial_r |LG_{\ell}(r,\phi)|^2 \overset{!}{=} 0.
\label{rmaxLG}
\end{align}
This leads to $r_{\textnormal{max}}=w_0\sqrt{\frac{\ell}{2}}$, which means that the intensity maxima of LG modes scale with $r_{\textnormal{max}}\propto \sqrt{\ell}$.

\subsection{Intuitive Explanation of the Radial Scaling}
We now give a more intuitive explanation of the radial scaling. This is important because the scaling is a key motivation for the rest of the article, as it leads to the unphysical prediction, which we later show how to solve in three entirely different physical system.
The square-root scaling can be understood in a more intuitive way by using a coordinate transformation which has been explicitly shown by Steuernagel \cite{steuernagel2005equivalence, steuernagel2010equivalence} (and in a more general context by Takagi \cite{takagi1990equivalence}): The idea is to rescale the transverse coordinates $x$ and $y$ in order make the intensity distribution of LG modes in eq.(\ref{LGmode}) independent of the distance $z$ from the focus: 
\begin{align}
x=\xi \frac{w(z)}{\sqrt{2}},\\
y=\eta \frac{w(z)}{\sqrt{2}},\\
z=z_R \tan(\tau)
\end{align}
Using
\begin{align}
\psi(x,y,z)=\tilde{\psi}(\xi,\eta,\tau) \exp^{\frac{i k (x^2+y^2)}{R(z)}}/w(z),
\end{align}
one finds the remarkable result that the paraxial wave equation transformed into the time-dependent Schr\"odinger for the harmonic oscillator
\begin{align}
\left(-\frac{\partial^2}{\partial \xi^2} -\frac{\partial^2}{\partial \eta^2} + \xi^2 + \eta^2 \right)\tilde{\psi}(\xi,\eta,\tau) =  2 i \frac{\partial}{\partial \tau}\tilde{\psi}(\xi,\eta,\tau)
\end{align}

The energies corresponding to the harmonic oscillator is $E_n=(n+1/2)\hbar$, while the potential scales with $V(x)=\alpha x^2$. Thus, by equating the energies, one sees that $x \propto \sqrt{n}$. Solutions to the harmonic oscillator are usually denoted as Hermite-Gauss functions in Cartesian coordinates. Laguerre-Gauss modes can be written as superpositions of Hermite-Gauss modes. The LG mode and the HG modes must have the same mode order, because the Gouy phase needs to match (see \cite{kimel1993relations} for a detailed decomposition of LG modes in terms of HG modes). As solutions of the harmonic oscillator scales with $x \propto \sqrt{n}$, and LG beams can be written as superpositions of them, also LG modes need to scale with $r_{\textnormal{max}} \propto \sqrt{\ell}$.

\subsection{Apparently unphysical Prediction}
One interesting prediction of these solutions is the following: The superposition of two beams with opposite OAM $\psi(\mathbf{r})=LG_{+\ell}(\mathbf{r})+LG_{-\ell}(\mathbf{r})$ has 2$\ell$ intensity maxima and minima at the radius $r=w_0 \sqrt{\frac{\ell}{2}}$. The azimuthal distance $\Delta$ between two intensity maxima and minima is given by
\begin{align}
\Delta&=\frac{2\pi r}{2\ell}=\frac{\pi w_0}{\sqrt{2\ell}}.
\end{align}
It means that the distance between two maxima decreases unboundedly as $\ell$ increases \cite{padgett2015divergence}, even to unphysically small values beyond the Rayleigh criterion\footnote{\textit{Superoscillations} allow for situations where high frequency oscillations with perfect visibilities can be achieved, in places where the intensity is exponentially small \cite{berry2006evolution}. However, we investigate situations where the field intensities are maximal.} or even beyond the Planck length. This is obviously a dubious situation, which motivates a more detailed investigation.

\subsection{What is done here?}
In our previous investigation of this effect \cite{krenn2016small}, we have used full Maxwell's solutions to solve this unphysical situation: The perfect intensity minima in the paraxial case are filled in by additional intensity contribution due to an field in the direction of propagation, which is neglected in the paraxial approximation. Here we show that the phenomenon happens not only for photons (massless Spin 1 particles) but also both for relativistic electrons (massive Spin 1/2 particles) and gravitational waves (massless Spin 2 field).

To do this, we use a method demonstrated by Iwo and Zofia Bialynicki-Birula for all of these particles and fields: A scalar potential function $\chi(\mathbf{r},t)$ which fulfills the D'Alambert wave equation is used to define the principle properties of the field (such as the helical phase pattern which corresponds to OAM or topological charge). Then three different vectorized generating function use the potential $\chi(\mathbf{r},t)$ to produce the vector field components for the electromagnetic fields for light \cite{bialynicki2006beams,bialynicki2013role}, for the Dirac spinor for electron \cite{bialynicki2017relativistic}, and for the Riemann curvature tensor for gravitational waves \cite{bialynicki2016gravitational}. 

In the non-paraxial and nonrelativistic limit, these solutions reassemble properties of the scalar solutions which are well studied mathematically and theoretically. In the non-paraxial and relativistic case, the method leads to full, non-approximative solutions of the different wave equations\footnote{In order to produce freely propagating solutions instead of evanescent waves, one needs to fullfil the criterion $k=\sqrt{k_x^2+k_y^2+k_z^2}=\frac{2\pi}{\lambda}$with $k_z \in \mathbb{R}$ .}. There are several ways how one could generalize these solutions into the non-paraxial regime. Therefore, we have compared the properties studied here with independent methods: For light, we have numerically applied the aplanatic lense model \cite{richards1959electromagnetic, novotny2012principles, fernandez2012helicity}, which is a standard way to describe focused fields (details can be found in \cite{krenn2016small}). For electrons, we have used the analytical solutions developed by Bliokh and others \cite{bliokh2011relativistic, bliokh2017theory} where the Spin-Orbit coupling can be conveniently seen. With all these methods, we find the same predictions for the properties we are studying, which is an important consistency check.

We thus see a very similar physical effect, which is encoded in three entirely different physical systems.

\section{Photons with Large Topological Charge}
\begin{figure}[t]
\includegraphics[width=0.5 \textwidth]{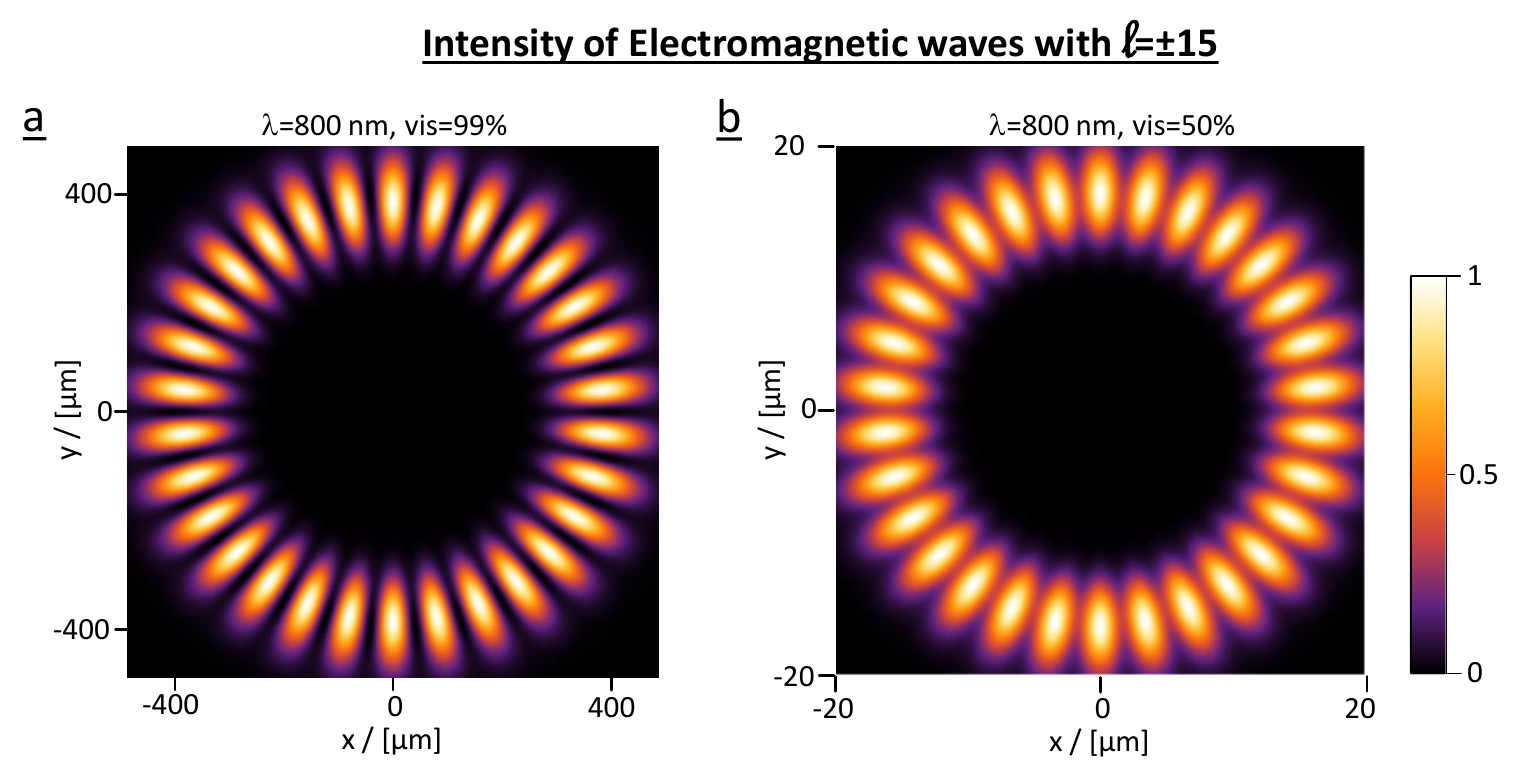}
\caption{Intensity of electromagnetic waves from eq.(\ref{eq:intensity_em}) with an OAM superposition of $\ell=\pm15 \hbar$ with visibilities $vis=0.99$ and $vis=0.5$, according to eq.(\ref{eq:visibility}). \textbf{a}: Here, the beam waist $w_0=149 \mu m$ is much larger than the wave length $\lambda=800 nm$. The minima and maxima are clearly visible. \textbf{b}: For a focused beam with $w_0=6.2 \mu m$ , the contrast between minima and maxima becomes smaller due to a growing field component $E_z$, whose intensity structure is shifted. A gaussian beam with a waist of $w_0=6.2 \mu m$ has an NA$\simeq \frac{\lambda}{\pi w_0}$=0.04. An investigation using the aplanatic lens model we have presented in \cite{krenn2016small}.}  
\label{fig:LightLGpm15}
\end{figure}
The paraxial wave equation is an approximation of the Maxwell equation for beams which are large compared to the wavelength \cite{lax1975maxwell, davis1979theory}. In order to describe strongly focused light beams, full solutions of the Maxwell's equations need to be used. In \cite{krenn2016small}, we have used the aplanatic lens model \cite{richards1959electromagnetic} to calculate electromagnetic fields after strong focus.

Here we use the elegant method of the Riemann-Silberstein vector \cite{silberstein1907elektromagnetische, bialynicki2013role}, which combines electric and magnetic fields in the form of
\begin{eqnarray}
\mathbf{F}=\frac{\mathbf{E}}{\sqrt{2 \epsilon_0}} + i \frac{\mathbf{B}}{\sqrt{2 \mu_0}}
\label{eq:RSVF}
\end{eqnarray}
where \textbf{F} is the Riemann-Silberstein vector, and \textbf{E} and \textbf{B} is the electric and magnetic field. The Riemann-Silberstein vector can be calculated using a generating function from a scalar potential $\chi(\mathbf{r},t)$ (which satisfies the d'Alambert differential equation). For a beam propagating in z-direction, one finds
\begin{eqnarray}
&\mathbf{F}=
\left(    
 \begin{array}{c}
      \partial_x \partial_z + \frac{i}{c} \partial_y \partial_t \\
      \partial_y \partial_z - \frac{i}{c} \partial_x \partial_t \\
      -(\partial_x^2 + \partial_y^2)
    \end{array} \right) 
    \chi(\mathbf{r},t)=\nonumber\\ 
&\left(    
 \begin{array}{c}
      \partial_- \partial_z + \frac{i}{c} \partial_+ \partial_t \\
      \partial_+ \partial_z - \frac{i}{c} \partial_- \partial_t \\
      -\left(\partial^2_r + \frac{1}{r}\partial_r + \frac{1}{r^2}\partial^2_{\phi}\right)
    \end{array} \right) 
    \chi(\mathbf{r},t)\nonumber\\  
\label{eq:RSVasPartial}
\end{eqnarray}
with $\mathbf{r}=(r,\phi,z)$, $\partial_+=\left(\sin\phi \partial_r + \frac{\cos\phi}{r} \partial_{\phi}\right)$ and $\partial_-=\left(\cos\phi \partial_r - \frac{\sin\phi}{r} \partial_{\phi}\right)$, and $\mathbf{E}=\sqrt{2\epsilon_0}Re(\mathbf{F})$ and $\mathbf{B}=\sqrt{2\mu_0}Im(\mathbf{F})$. For Laguerre-Gaussian beams, the potential function can be written as \cite{bialynicki2006beams}

\begin{eqnarray}
\chi_{\Omega,\ell}^{\sigma}(\mathbf{r},t)=N\cdot \frac{e^{-i\sigma\left(\Omega \left(t - \frac{z}{c}\right)-\ell\phi \right)}r^{|\ell|}}{a(t,z)^{|\ell|+1}} e^{-\frac{r^2}{a(t,z)}}
\label{eq:RSVpotentialLG}
\end{eqnarray} 
with $a(t,z)=w_0^2 + i\frac{\sigma c^2 \left(t+\frac{z}{c}\right)}{\Omega}$, where $N$ is a normalisation constant, $\sigma =\pm1$ is the circular polarisation (we use right-handed circular polarisation $\sigma=1$ in the whole manuscript), $\Omega=\frac{c}{\lambda}$ is the optical (mean) frequency and $w_0$ is the beam waist at the focus. 

For large beam waists $w_0$, the fields reassemble the property of the scalar Laguerre-Gauss beams  in eq.(\ref{LGmode}). For example, when one creates superpositions of positive and negative topological charges
\begin{eqnarray}
\chi_{\Omega,\pm\ell}^{\sigma}(r,\phi,z,t)=\frac{1}{\sqrt{2}}\left(\chi_{\Omega,+\ell}^{\sigma}(\mathbf{r},t)+\chi_{\Omega,-\ell}^{\sigma}(\mathbf{r},t)\right),\nonumber\\
\label{eq:RSVSuperpos}
\end{eqnarray} 
one finds an electromagnetic intensity distribution
\begin{eqnarray}
\textnormal{I}(r,\phi)=|\mathbf{F}|^2
\label{eq:intensity_em}
\end{eqnarray}
with 2$\ell$ maxima and minima distributed on a ring, as seen in Fig.\ref{fig:LightLGpm15}a. Interestingly, for strongly focused beams and large values of $\ell$, the intensity in the minima increases (starting from zero intensity in the paraxial case), as can be seen in Fig.\ref{fig:LightLGpm15}b. This is because of an increasing electromagnetic field component E$_z$ and B$_z$, which is neglected in the scalar solution in eq.(\ref{LGmode}) \footnote{The existence of a longitudinal component in the electromagnet field follows directly from Gauss' law of Maxwell's equations. For a uniformly polarized beam, it is easy to calculate the longitudinal $E_z$ and $B_z$ components given the transverse components. A very good summary of this method is shown in \cite{carnicer2012longitudinal}.}. The additional intensity contribution has been described and discussed in the context of the diffraction limit in \cite{krenn2016small}, and has recently been observed experimentally \cite{wozniak2016tighter}. 

From the intensity $\textnormal{I}(r,\phi)$, one can calculate the visibility between maxima and minima for the radius of maximum intensity. The radius of maximal intensity can be obtained (analog as for the paraxial case in eq.(\ref{rmaxLG})) as $\partial_r |\textnormal{I}(r,\phi_0)|^2 \overset{!}{=} 0$ which leads to $r_{max}$ (with $\phi_0$ being the angle where the intensities constructively interfere at the ring). The intensity in azimuthal direction $\textnormal{I}(r_{max},\phi))$ has maxima and minima,
\begin{eqnarray}
\textnormal{I}_{\textnormal{max}}=\underset{\phi}{\textnormal{max}}\big(\textnormal{I}({r_{max}},\phi)\big)\nonumber\\
\textnormal{I}_{\textnormal{min}}=\underset{\phi}{\textnormal{min}}\big(\textnormal{I}({r_{max}},\phi)\big).
\label{eq:minima}
\end{eqnarray} 
To quantify the modulation, we define the visibility as 
\begin{eqnarray}
vis=\frac{\textnormal{I}_{\textnormal{max}}-\textnormal{I}_{\textnormal{min}}}{\textnormal{I}_{\textnormal{max}}+\textnormal{I}_{\textnormal{min}}}
\label{eq:visibility}
\end{eqnarray} 

\subsection{Why is the additional field out of phase?}
The contrast between the minima and maxima of the intensity distribution decreases for smaller beam, because the additional field in z-direction $F_z$ (the longitudinal component) has an amplitude which is shifted with respect to the transverse components $F_x$ and $F_y$ (i.e. the maxima of $F_z$ are located at the minima of $F_x$ and $F_y$). This can be understood by observing that the potential for the superposition in eq.(\ref{eq:RSVSuperpos}) can be written as $\chi_{\Omega,\pm\ell}^{\sigma}(r,\phi,z,t)=\tilde{\chi}_{\Omega}^{\sigma}(r,z,t)\cdot\sin\left(\ell \phi \right)$. From eq.(\ref{eq:RSVasPartial}) one can see that the dominant term in $F_x$ and $F_y$ scales with $\mathcal{O}\left(\frac{\partial_{\phi}}{r}\right)=\mathcal{O}\left(\frac{\ell}{r}\right)$ while the dominant term for $F_z$ scales with $\mathcal{O}\left(\frac{\partial_{\phi}^2}{r^2}\right)=\mathcal{O}\left(\frac{\ell^2}{r^2}\right)$. Thus, when $\ell$ is increased or the beam waist $w_0$ (and thereby $r$) decreases by stronger focusing, the non-paraxial propagation starts dominating and forming the longitudinal field component $F_z$. Furthermore, the two competing terms generated by a first and second derivative of a $\sin(\ell \phi)$, thus one term is $\cos(\ell \phi)$ while the other is $\sin(\ell \phi)$ -- and they are out of phase. This is the reason why the transverse components $F_x$ and $F_y$ are out of phase with the longitudinal component $F_z$.

These findings are relevant in quantum entanglement experiments as well: The intensity structures have been used to verify entanglement between photons with very large values of angular momentum, such as $\ell=$300$\hbar$ \cite{fickler2012quantum} and recently (with the help of spiral phase mirrors) with $\ell=$10.000$\hbar$ \cite{fickler2016quantum}. If such entangled photons were focused, the entanglement cannot be measured with simple intensity masking anymore.

\section{Relativistic Electrons with Large Topological Charge}
\begin{figure}[b]
\includegraphics[width=0.5 \textwidth]{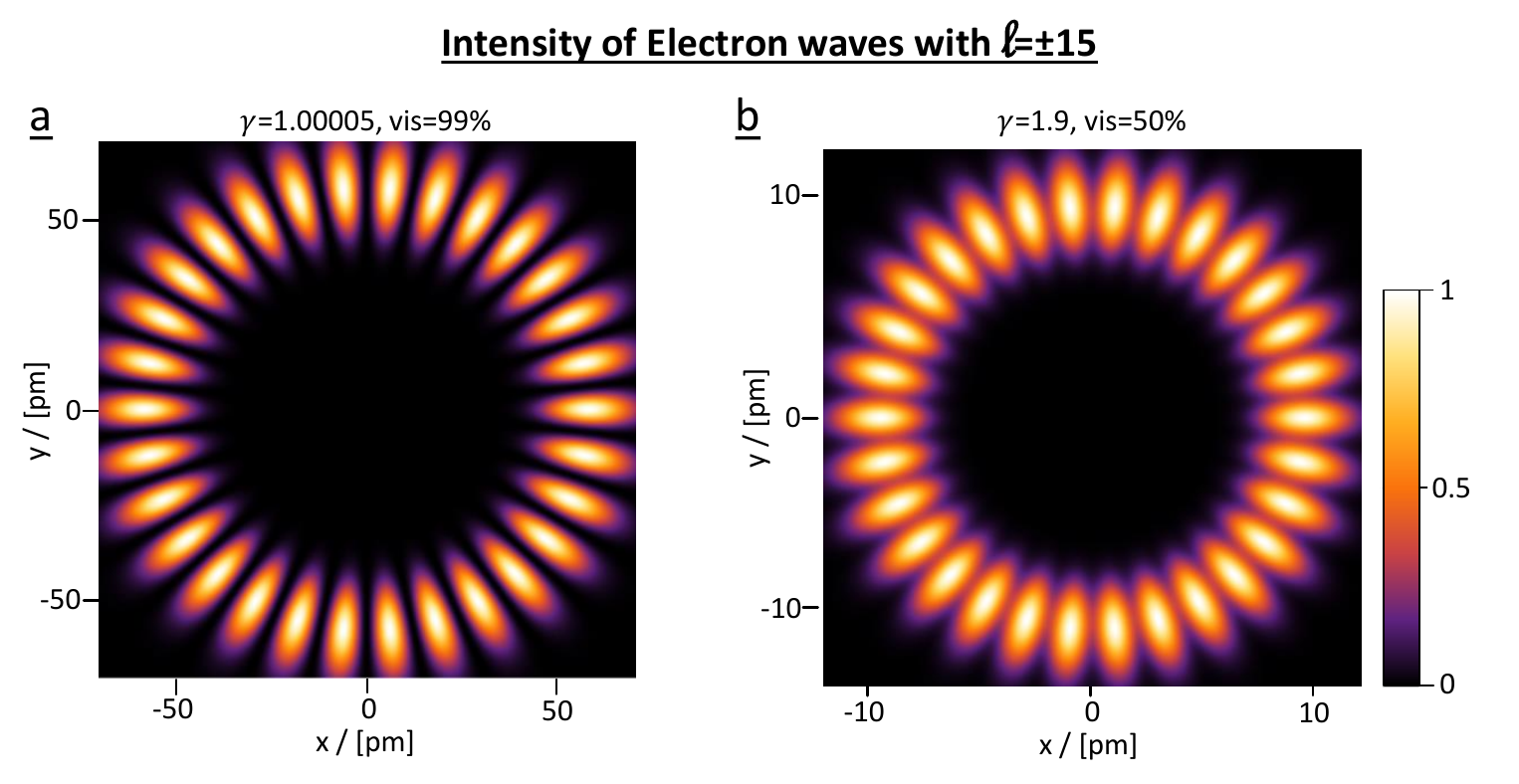}
\caption{Intensity of electron beams from eq.(\ref{eq:intensityElectron}) with an OAM superposition of $\ell=\pm15 \hbar$, with visibilities according $vis=0.99$ and $vis=0.5$, to eq.(\ref{eq:visibility}). \textbf{a}: A nonrelativistic electron beam (with $b=1500$, and the relativistic Lorentz factor $\gamma=1+5\cdot 10^{-5}$) has clearly visible minima and maxima. \textbf{b}: A focused relativistic e$^-$ beam (with $b=142.1$, and the relativistic Lorentz factor $\gamma=1.9$) shows that the contrast between minima and maxima decreases.}  
\label{fig:e-LGpm15}
\end{figure}

For analysing the energy distribution of electrons with superposed OAM beams, we use the recently derived analytic and finite-energy solution in \cite{bialynicki2017relativistic}, which derives the Dirac spinor for the relativistic electron using a generating function:

\begin{eqnarray}
&f_{\ell}(\mathbf{r},t)=N\cdot \exp\left(i p_z z / \hbar\right) \nonumber\\
&\cdot\exp\left(i \ell \phi\right) \frac{\exp\left(-b h(r,t)\right)}{h(r,t)} \left(\frac{q r}{h(r,t) + 1 + i q c t} \right)^{|\ell|}
\label{eq:electronPotential}
\end{eqnarray} 
with $h(r,t)=\sqrt{(1+i q c t)^2 + (q r)^2}$, b is dimensionless and describes the width of the wave packet, $q=\gamma / b \lambdabar$, $\gamma=\sqrt{1+(p_z/mc)^2}$ is the relativistic factor, and $\lambdabar=\hbar / m_e c = 3.86 \cdot 10^{-13}$ m is the reduced Compton wavelength, with the mass of the electron $m_e$.

To obtain the Dirac spinor $\Psi(\mathbf{r},t)$ for the electron, one can use the generating function

\begin{eqnarray}
\Psi_{\ell}(\mathbf{r},t)=
\left(    
 \begin{array}{c}
      1 \\
      0 \\
      i \lambdabar
    \left(    
     \begin{array}{c}
          \frac{1}{c} \partial_t + \partial_z \\
          e^{i\phi}(\partial_r + \frac{i}{r}\partial_{\phi}) 
    \end{array}
    \right)
\end{array}
\right) 
    f_{\ell}(\mathbf{r},t).  
\label{eq:electronDiracSpinor}
\end{eqnarray}

Electrons with a superposition of opposite OAM are described by
\begin{eqnarray}
\Psi_{\pm\ell}(\mathbf{r},t)=\frac{1}{\sqrt{2}}\left(\Psi_{+\ell}(\mathbf{r},t) + \Psi_{-\ell}(\mathbf{r},t)\right).   
\label{eq:DiracSpinorSuperpos}
\end{eqnarray}

The probability density
\begin{eqnarray}
\textnormal{I}(r,\phi)=\Psi^{\dagger}\Psi.   
\label{eq:intensityElectron}
\end{eqnarray}
for a nonrelativistic and relativistic case is shown in Fig.\ref{fig:e-LGpm15}. It clearly shows that the contrast between minima and maxima decreases in a similar way as they did for photons. In contrast to the photonic case where an additional field in the direction of propagation emerged, the reason for electrons is different: The additional probability arises due to a large contribution the fourth component of $\Psi_{\ell}(\mathbf{r},t)$. That component stands for a $S=-\frac{1}{2}$ contribution in the rest frame of the electron. It represents a \textit{spin-orbit coupling for relativistic electrons} \cite{bliokh2017theory}. Its minima are at the position of the maxima of first component of $\Psi_{\ell}(\mathbf{r},t)$, which is the leading contribution for the nonrelativistic case.

An analogous effect occurs for relativistic electrons in Bessel modes, which have been described for the first time in 2011 \cite{bliokh2011relativistic, bliokh2017theory} (also described in \cite{bialynicki2017relativistic}). Bessel beams are not normalizable therefore would require an infinite amount of energy for their generation. For this reason, we describe the situation with finite-energy Laguerre-Gaussian beams.

The investigation above could be interesting because powerful experimental methods have been developed in recent year for free electrons carrying orbital angular momentum (see for example here: \cite{bliokh2017theory, grillo2016generation}) since they have been observed for the first time in 2010 \cite{uchida2010generation, verbeeck2010production}. In particular, electrons with up to 100$\hbar$ \cite{mcmorran2011electron}, 200$\hbar$ \cite{grillo2015holographic} and 1000$\hbar$ \cite{mafakheri2017realization} of orbital angular momentum have been created recently. It would be interesting whether this effect can be observed in experiments.

\section{Gravitational Waves with Large Topological Charge}
\begin{figure}[b]
\includegraphics[width=0.5 \textwidth]{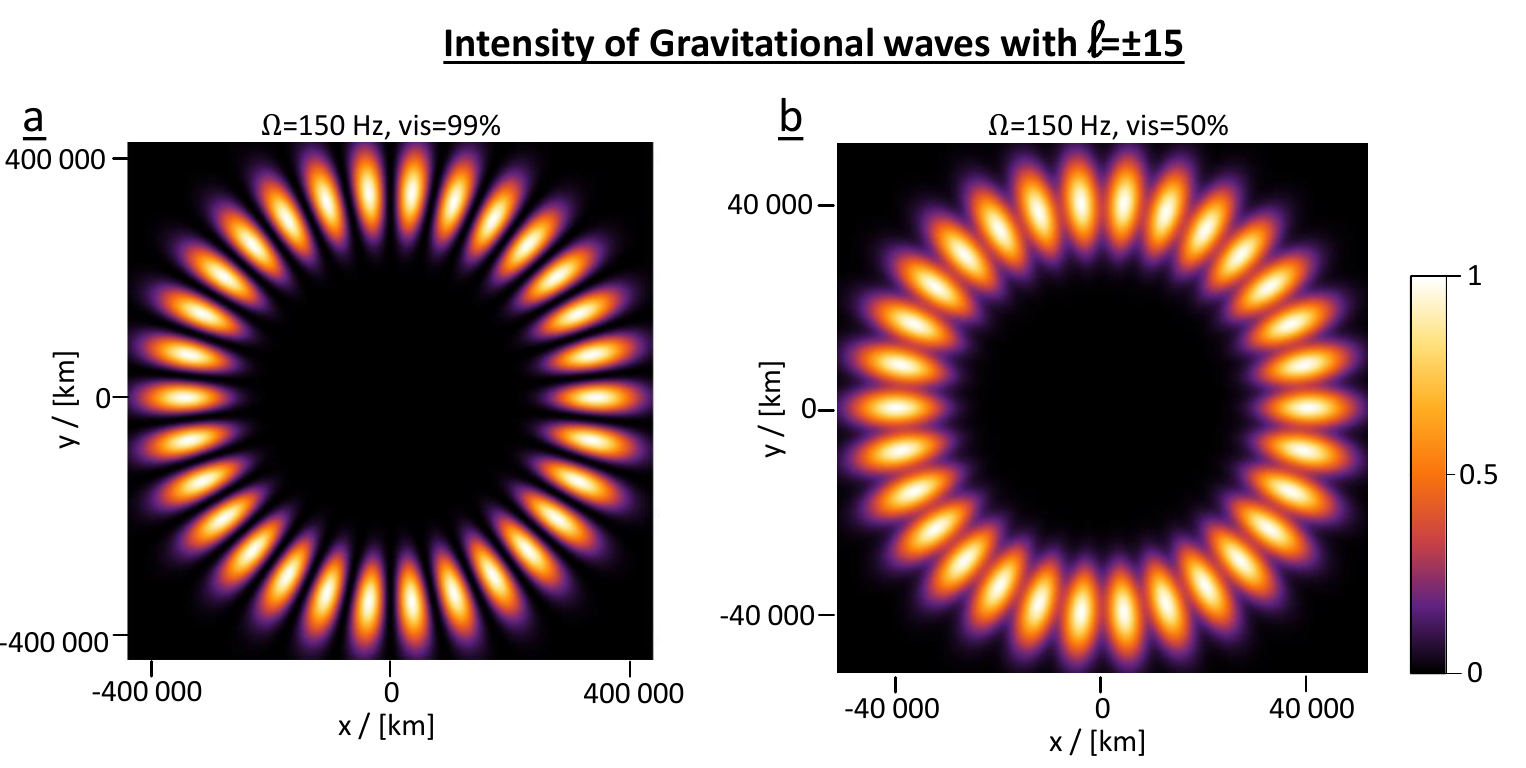}
\caption{Intensity of Gravitational Wave from eq.(\ref{eq:intensityGW}) with an OAM superposition of $\ell=\pm15 \hbar$, with visibilities $vis=0.99$ and $vis=0.5$, according to eq.(\ref{eq:visibility}). \textbf{a}: A large gravitational wave (with with a frequency $\Omega=150Hz$ similar as GW150914 \cite{abbott2016observation} and $w_0=63.1\lambda$. \textbf{b}: A focused gravitational wave (with $w_0=7.41\lambda$) with 50\% visibility.}  
\label{fig:graviLGpm15}
\end{figure}
Gravitational waves in free space can be described by linearized gravity, which is a weak-field solution of Einstein's General Relativity (GR) in the absence of matter. The similarities between electromagnetic and gravitational waves has been used to employ well-developed techniques from electromagnetism to study properties of weak-field GR (see for instance a recent and very illuminating article by Barnett \cite{barnett2014maxwellian} and references therein). In particular, the construction of solutions of gravitational waves allows for the usage of potentials and generating functions \cite{penrose1965zero, stewart1979hertz}. Based on a method by Penrose, Iwo and Zofia Bialynicki-Birula have shown a technique to construct analytic solutions for gravitational waves carrying OAM \cite{bialynicki2016gravitational}. We use their technique to investigate focussed gravitational waves with a superposition of $\pm\ell$.  Analog to the electromagnetic case in eq.(\ref{eq:RSVF}), one constructs a self-dual version of the Riemann curvature tensor
\begin{eqnarray}
G_{\mu\nu\lambda\rho}=R_{\mu\nu\lambda\rho} + \frac{i}{2} \epsilon_{\mu\nu\alpha\beta} R_{\lambda\rho}^{\alpha\beta}  
\label{eq:graviDualTensor}
\end{eqnarray}
and in analogy to the Riemann-Silberstein vector in electrodynamics, one defines
\begin{eqnarray}
\mathcal{G}_{ij}=G_{0i0j}
\label{eq:graviGij}
\end{eqnarray}
which is a symmetric 3x3 tensor that carries the full information about the linearized gravitational field. It has been extensively studied in electromagnetic analogies for weak-field gravity \cite{maartens1998gravito}. The six different components can be obtained via a potential function using 
\begin{eqnarray}
\left(    
 \begin{array}{c}
      \mathcal{G}_{11} \\
      \mathcal{G}_{12} \\
      \mathcal{G}_{13} \\
      \mathcal{G}_{22} \\
      \mathcal{G}_{23} \\
      \mathcal{G}_{33} \\
\end{array}
\right) 
=
\left(
 \begin{array}{c}
      \phi_{0000} - 2 \phi_{0011} + \phi_{1111} \\
      i\phi_{0000} - i\phi_{1111} \\
      2\phi_{0111} - 2\phi_{0001} \\
      -\phi_{0000} - 2 \phi_{0011} - \phi_{1111} \\
      -2i\phi_{0001} - 2i\phi_{0111} \\
      4 \phi_{0011} \\
\end{array}
\right)
\label{eq:GfullEq}
\end{eqnarray}
with
\begin{eqnarray}
\phi_{ABCD}=\left(\frac{1}{c}\partial_t - \partial_z \right)^l \left(-e^{i\phi}\left(\partial_r + \frac{i}{r} \partial_{\phi}\right)\right)^n \chi(\mathbf{r},t)\nonumber\\
\label{eq:electronDiracSpinor}
\end{eqnarray}
where $l$ and $n$ are the numbers of zeros and ones in $ABCD$, respectively. Same as in the electromagnetic and electronic case, $\chi(\mathbf{r},t)$ is the potential function which is a solution of d'Alembert's equation. Analog to the electromagnetic case, the intensity distribution can be obtained by 
\begin{eqnarray}
\textnormal{I}(r,\phi)=\frac{1}{2}\mathcal{G}^{ij}\mathcal{G}^*_{ij}.
\label{eq:intensityGW}
\end{eqnarray}
We use the same potential function as in the electromagnetic case in eq.(\ref{eq:RSVSuperpos}) and derive the solution for the gravitational wave carrying OAM with eq.(\ref{eq:GfullEq}), using $\ell=15$ and the mean frequency $\Omega$=150 Hz, which was the frequency of the first observed gravitational wave \cite{abbott2016observation}. In Fig.\ref{fig:graviLGpm15} it is shown that for very large gravitational waves, the intensity distribution has the expected structure with 30 minima and maxima aligned in a ring. However, when the gravitational wave is much smaller, the minima are filled and the visibility between minima and maxima drops significantly. 

The origin of the additional contributions are components $\mathcal{G}_{13}$ and $\mathcal{G}_{23}$, which can not be neglected anymore when the gaussian beam waist is small and $\ell$ is large. These components are connected with the curvature of the space-time between transversal and longitudinal direction in the Riemann curvature tensor.    

\section{Discussion}
The paraxial, small-angle solutions for beams carrying orbital angular momentum predicts a curious, unphysical effect which would allow for breaking the Rayleigh criterion. Recently we have shown for beams of light, using full Maxwell's equation, that this prediction is eliminated by an additional field in the direction of propagation \cite{krenn2016small}, which is neglected in the paraxial approximation. Here we analysed the predictions for electrons and gravitational waves carrying orbital angular momentum. This was possible using a unified method based on scalar potentials and generating functions developed recently by Iwo and Zofia Bialynicki-Birula. We find that in both systems, using full Dirac equations for free electrons and linearized general relativity for gravitation waves, additional intensity distributions emerge -- in a very similar way as for the photonic case, but for different physical reasons. In electrons, a relativistic spin-to-orbit coupling accounts for the modified intensity distribution while for gravitational waves, curvature between the transverse and longitudinal direction of space-time contributes additional intensity.

On the experimental side, this effect has been demonstrated in an impressive experiment for photons \cite{wozniak2016tighter}. It would be very interesting whether it can also be experimentally confirmed with other particles such as electrons with large OAM \cite{grillo2015holographic, mafakheri2017realization}.

Also, several questions remain open: How can these phenomena be explained intuitively? Can this effect be explained without taking advantage of the specific physical implementation? Is there a reason stemming from the uncertainty principle which can explain the predicted behaviour? \footnote{Ole Steuernagel points out that ref.\cite{feng2001physical} (which deals with the uncertainty relation for higher-order spatial modes) could be relevant for this question.}\footnote{Dmitry Karlovets points out that when one focuses tightly a superposition of beams, the Wigner function of such a state acquires regions of negativity in phase phase, as the beam starts to interfere with itself in free space \cite{karlovets2017possibility}.}

What happens to other types of strongly focussed particles carrying large topological charge\footnote{Boris Malomed points out that a polariton-like model of a spinor BEC admits stable 2D bright-soliton solutions with arbitrarily high values of embedded vorticity \cite{qin2016stable}. It might be interesting to investigate such systems as well.}, such as molecular matter waves which can be composed of 100s of atoms \cite{arndt1999wave, eibenberger2013matter}; free propagating neutrons with OAM \cite{clark2015controlling, larocque2018twisting} -- especially when they decay; neutrinos -- especially as their flavors oscillate \cite{fukuda1998evidence,ahmad2002direct}; or to fundamental scalar particles \cite{aad2012observation, chatrchyan2012observation} as vectorial characters of the field (which are responsible for the decreasing contrast between minima and maxima of the particles investigated here) might not contribute to the intensity distribution?

We have to come back to these questions another time.

\section*{Acknowledgements}
The authors thank Konstantin Bliokh, Ebrahim Karimi and Vincenzo Grillo for interesting discussions on electrons with OAM, and Iwo and Zofia Bialynicki-Birula, Ole Steuernagel and Robert Fickler for useful comments on the manuscript. This work was supported by the Austrian Academy of Sciences (\"OAW), and the Austrian Science Fund (FWF) with SFB F40 (FOQUS).

\bibliographystyle{unsrt}
\bibliography{refs}

\end{document}